\documentclass[a4paper,11pt]{article}
\usepackage{pos}

\title{Jet quenching in the hadron gas: An exploratory study}
\ShortTitle{Hadronic jet quenching}

\author*[a,b,c]{H. Elfner}
\author[b]{P. Dorau}
\author[b,c]{J.-B. Rose}
\author[d]{D. Pablos}

\affiliation[a]{GSI Helmholtzzentrum f\"ur Schwerionenforschung, \\
Planckstr. 1, 64291 Darmstadt, Germany}

\affiliation[b]{Institute for Theoretical Physics, Goethe University,\\
Max-von-Laue-Strasse 1, 60438 Frankfurt am Main, Germany}

\affiliation[c]{Frankfurt Institute for Advanced Studies, \\
Ruth-Moufang-Strasse 1, 60438 Frankfurt am Main, Germany}

\affiliation[d]{Department of Physics and Technology, University of Bergen, \\5007 Bergen, Norway}

\emailAdd{h.elfner@gsi.de}
\emailAdd{dorau@fias.uni-frankfurt.de}
\emailAdd{rose@fias.uni-frankfurt.de}
\emailAdd{Daniel.Pablos@uib.no}

\abstract{In most calculations of hard particle suppression in heavy-ion reactions the hadronic stage has been neglected due to formation time arguments. Most of the hard particle shower exits the hot and dense medium before the system enters the hadronic evolution. In this contribution, a first assessment within the hadronic transport approach SMASH (Simulating Many Accelerated Strongly-interacting Hadrons) of rescattering effects on hard particles is presented. In particular, it is shown that the hadronic energy loss depends on the particle species as well as the energy of the probe. A parametrization for the $\langle \tilde{q} \rangle$ parameter as a function of temperature and particle energy is given for pions. Overall, major effects of the hadronic stage are expected in the transverse momentum range from 2-10 GeV and therefore jet sub-structure analysis and (hard-soft) correlation observables might be affected.}

\FullConference{%
  HardProbes2020\\
  1-6 June 2020\\
  Austin, Texas}

\begin{document}
\maketitle

\section{Introduction}
Jets or showers of particles initiated by a hard parton are getting modified by the strongly-interacting medium created in ultra-relativistic heavy-ion collisions. In that sense they serve as a tomographic probe of the quark-gluon plasma \cite{Mehtar-Tani:2013pia,Qin:2015srf}. For any conclusion on medium properties and a detailed understanding of the energy loss mechanisms a sophisticated dynamical description of the medium is essential \cite{Bass:2008rv}. The standard picture for the evolution of heavy-ion reactions consists of an initial non-equilibrium evolution that serves as an initial state for 3+1-dimensional viscous hydrodynamic evolution followed by subsequent hadronic rescattering. Due to formation time arguments most of the calculations for hard probes only consider the interaction with the hydrodynamic evolution in the quark-gluon plasma phase and the hadronic effects have been neglected. This contribution summarizes the findings in \cite{Dorau:2019ozd} to demonstrate how hard particles will be affected during the late stage evolution. Other studies within hadronic transport approaches (HSD and UrQMD) indicate that there might be significant modification of the transverse momentum spectrum in the range of $p_T = 2-10$ GeV by hadronic scattering \cite{Cassing:2003sb,daSilva:2020cyn}. In this work, the hadronic transport approach SMASH-1.6 (Simulating Many Accelerated Strongly-interacting Hadrons) \cite{Weil:2016zrk,SMASH16} has been employed to assess the effects of hadronic processes on particles with high transverse momentum.

\section{Influence on jet shape observables}

\begin{figure}[h]
\centering
 \includegraphics[width=0.45\textwidth]{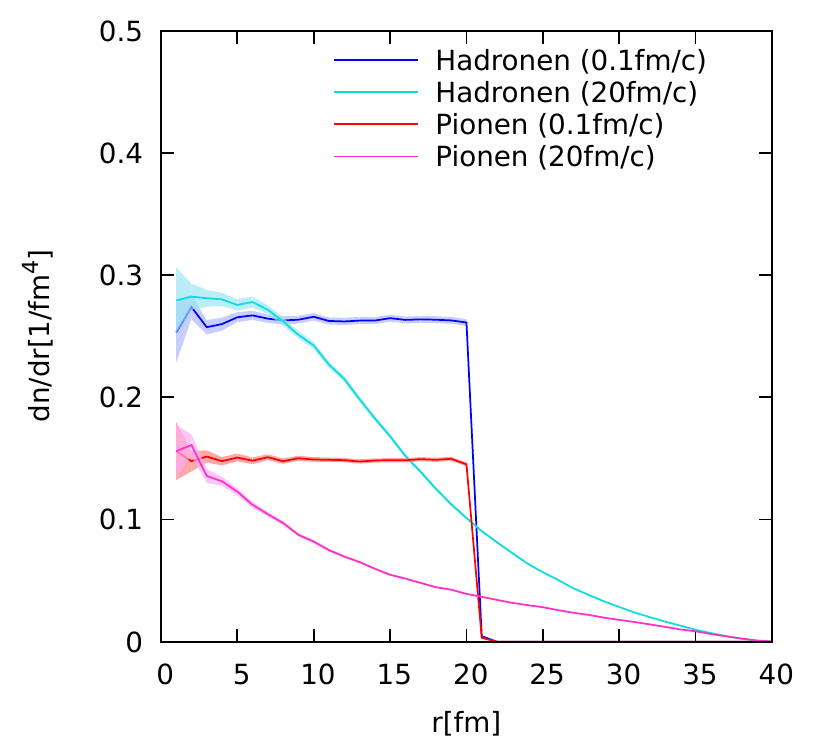}
 \includegraphics[width=0.42\textwidth]{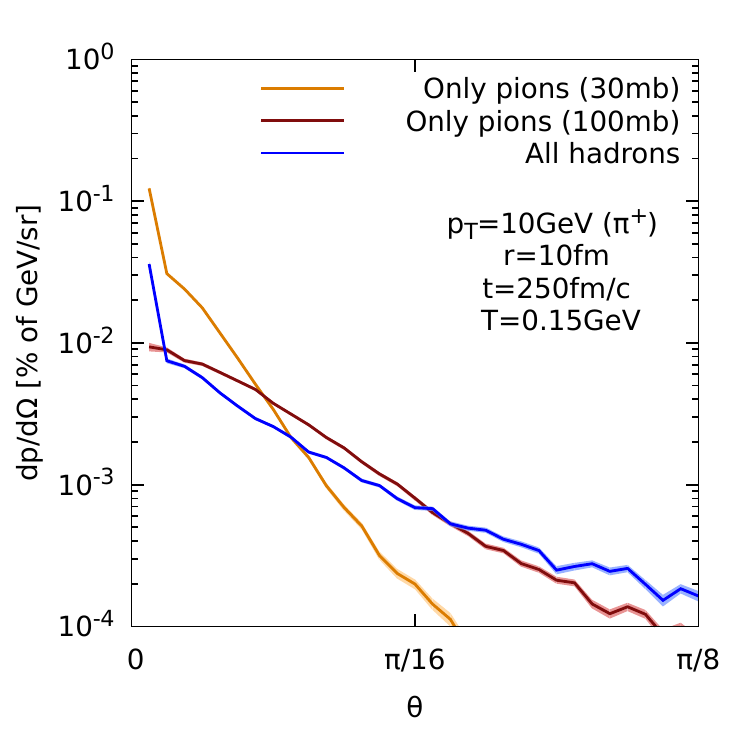}
\caption{Left: Distribution of particles (pions or all hadrons) in a sphere of radius $r=20$ fm at initial and final time of the evolution. Right: Comparison of jet shapes in an expanding sphere with only pions and fixed cross-section to the full hadron gas (taken from \cite{Dorau:2019ozd}).}
\label{fig_sphere}
\end{figure}

As a simplified scenario, an expanding sphere containing only pions or a full hadron gas is used. Fig. \ref{fig_sphere} (left) shows the initial uniform radial distribution of particles as well as the final distribution after the expansion.

The density of particles differs by a factor of two due to initialisation with multiplicities that correspond to a temperature of $T=150$ MeV which is a typical switching temperature in hybrid approaches. The spherical expansion is calculated once with the hard particle and once without it to allow for easy subtraction of background. In Fig. \ref{fig_sphere} (right) only the angular distribution associated with the hard particle is shown. For a pion with $p_T=10$ GeV there is significant broadening already for a small cross-section of $\sigma=10$ mb. The main interesting observation is that a pion gas with fixed cross-section of $\sigma=100$ mb corresponds roughly to a full hadron gas at the same temperature.

\section{Extraction of jet quenching parameters}

\begin{figure} [t]
\centering
 \includegraphics[width=0.45\textwidth]{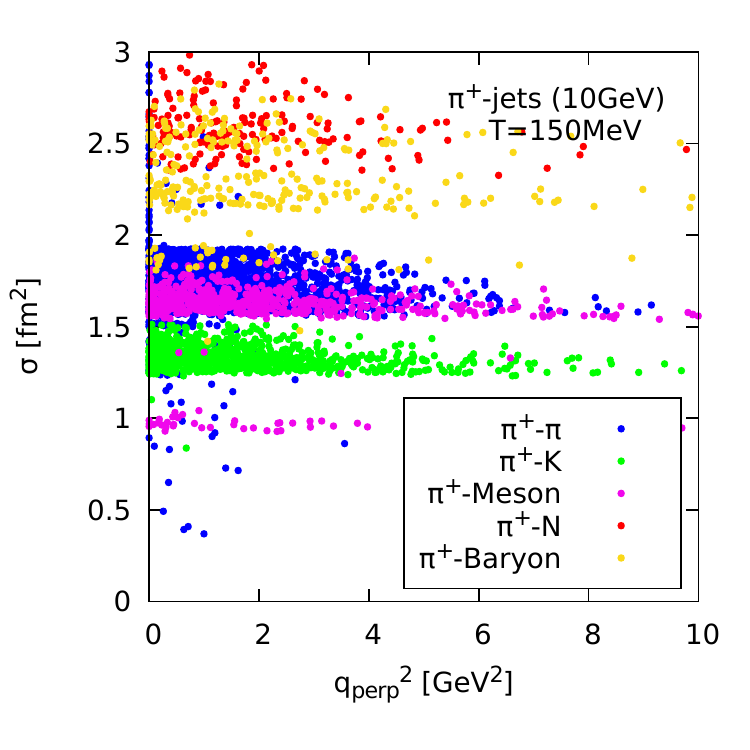}
 \includegraphics[width=0.44\textwidth]{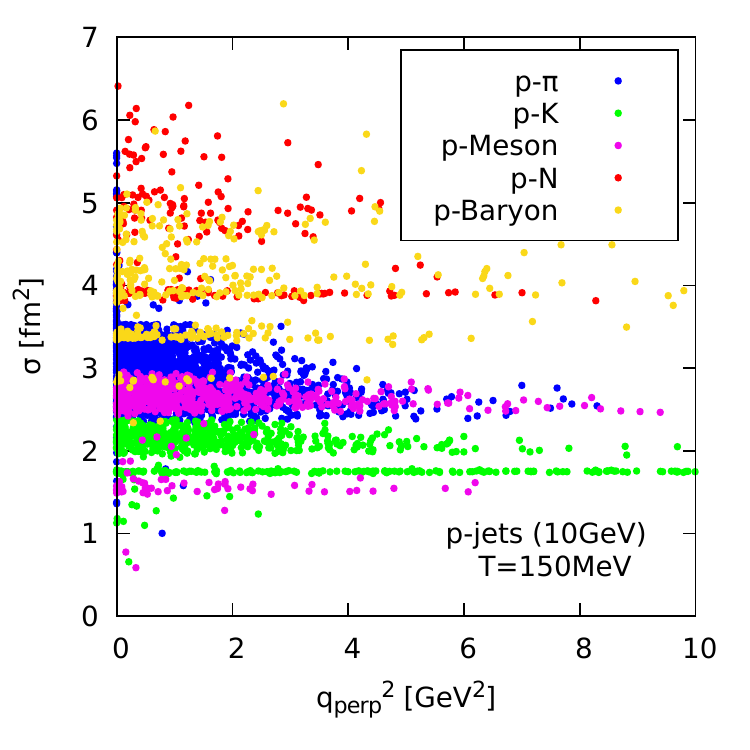}
\caption{Scatter plot of the cross-sections for the binary scatterings of pions (left) and protons (right) as a function of transverse momentum transfer.}
\label{fig_scatter}
\end{figure}

To extract the transport coefficient associated with jet quenching, namely the transverse momentum difference per unit length, $\hat{q}$, an infinite matter simulation in a box with periodic boundary conditions has been performed. A full hadron gas with all degrees of freedom available in SMASH has been generated at a temperature of $T=150$ MeV. Fig. \ref{fig_scatter} indicates the different types of processes that a pion (left) and a proton (right) with $p_T=10$ GeV encounter. In kinetic theory, $\hat{q}$ is calculated from a weighted integral of the differential cross-section as a function of transverse momentum transfer $\frac{d\sigma}{dq_\perp^2}$. This is impossible in our calculation, since only the size of the total cross-section decides over the probability for scatterings. There are different clusters of reaction types independent of $q_\perp$ and for protons and pions the reactions with other baryons are largest. 
\begin{figure} [t]
\centering
 \includegraphics[width=0.43\textwidth]{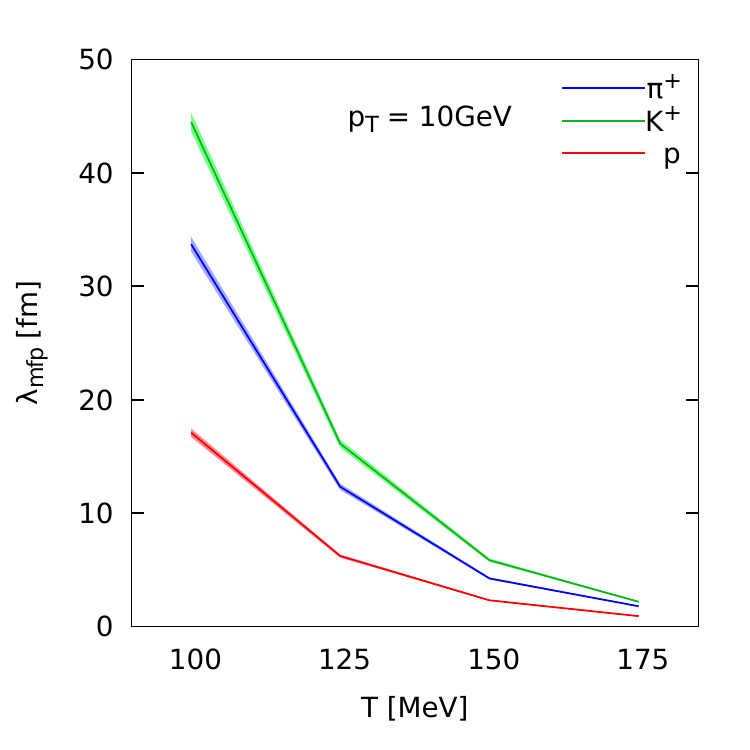}
 \includegraphics[width=0.43\textwidth]{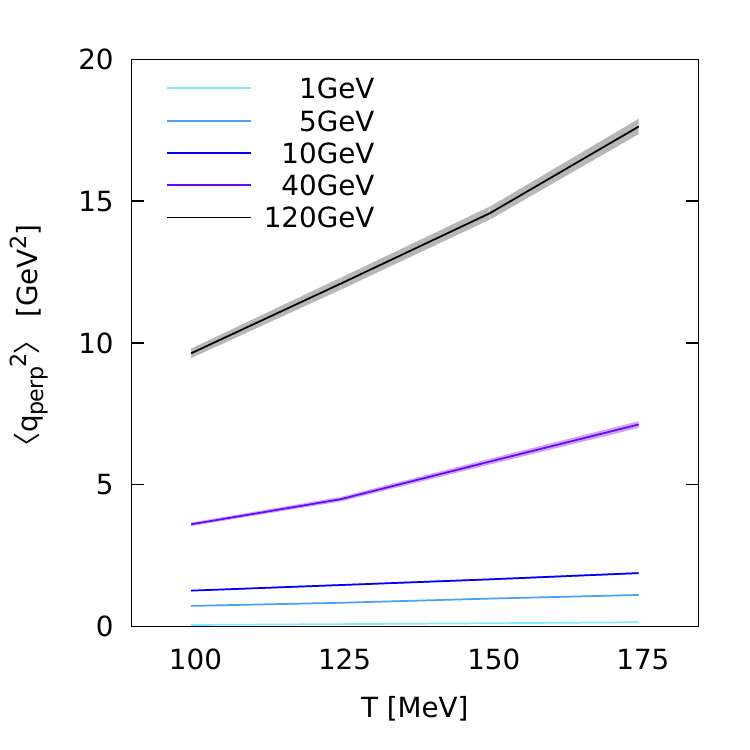}
\caption{Left: Mean free path of first collisions as a function of temperature for a pion with 10 GeV energy. Right: Squared average momentum transfer as a function of temperature for different probe energies (taken from \cite{Dorau:2019ozd}).}
\label{fig_lambdaqperp}
\end{figure}

The species dependence of the total cross-sections translates in correspondingly different mean distances to the first interaction shown in Fig. \ref{fig_lambdaqperp} (left). 
Since most of the reactions are inelastic, we cannot follow the particles through several interactions, but have to restrict our analysis to the first interaction that happens. Strange particles have in general smaller cross-sections and therefore the mean free path is largest for kaons. Fig. \ref{fig_lambdaqperp} (right) shows the temperature dependence of the average momentum transfer transverse to the direction of movement of the hard pion. The higher the initial energy of the probe the more momentum is available to be lost in the collisions.

\begin{figure} [h]
\centering
 \includegraphics[width=0.43\textwidth]{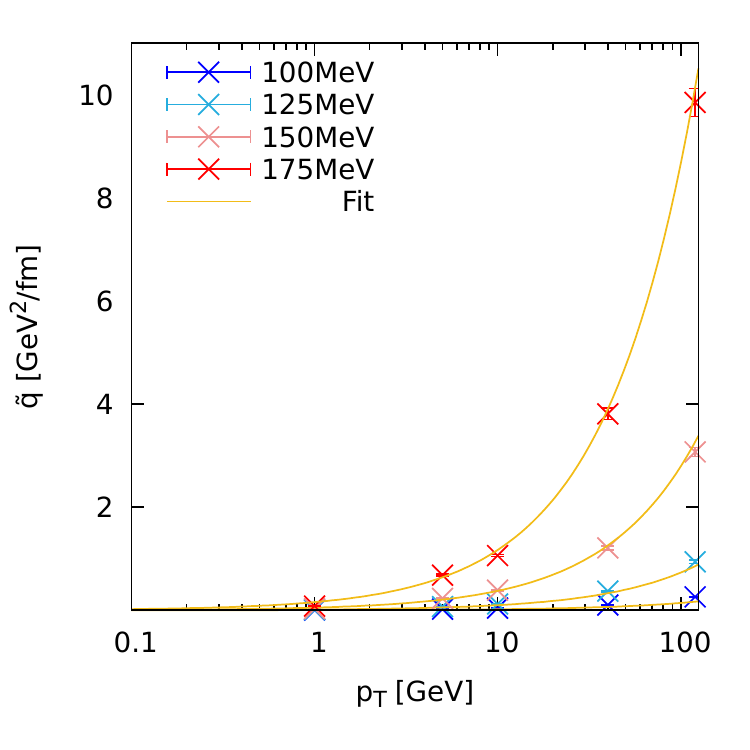}
 \includegraphics[width=0.43\textwidth]{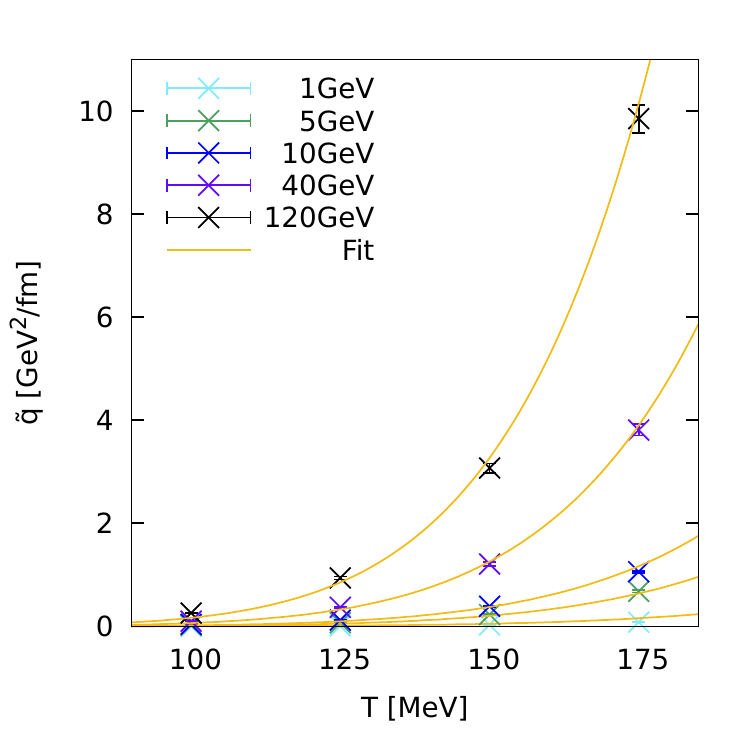}
\caption{Slices through the two-dimensional function of $\tilde{q}$ as a function of temperature  for different transverse momenta of the pion (taken from \cite{Dorau:2019ozd}).}
\label{fig_qtilde}
\end{figure}

Finally, by dividing the mean transverse momentum transfer by the mean free path results in the quantity
\begin{equation}
\tilde{q} = \frac{q_\perp^2}{\lambda_{\rm{mfp}}}
\end{equation}
as a coefficient that quantifies the hadronic energy loss. Fig. \ref{fig_qtilde} indicates how the temperature and probe energy dependence of $\tilde{q}$ can be parametrized. More details on the 2-dimensional parametrization as well as results for the longitudinal momentum transfer can be found in \cite{Dorau:2019ozd}. The extracted values are on the order of 20-30 \% of the typical values for $\hat{q}$ that are reported from partonic calculations.

\section{Conclusion and Outlook} 
The results presented here are ready to be used within a probabilistic energy loss approach in the hadronic stage, when it is described by hydrodynamics. In the future, it would be interesting to feed the hadrons from medium modified jet showers in the hadronic afterburner to investigate the consequences on observables. In general, the largest influence of hadronic rescatterings is expected in the transverse momentum range of $p_T=2-10$ GeV and therefore, substructure and correlation observables will be most affected. A major open question that is beyond the work presented here is how to properly address the interactions of partons with hadrons, when a parton propagates through a hadronic environment, which is also relevant for future EIC studies.   

\section*{Acknowledgements} 
This work was initiated during a visit in the context of the PPP exchange project with McGill University supported by the DAAD funded by BMBF (Project-ID 57314610). Computational resources have been provided by the Center
for Scientific Computing (CSC) at the Goethe-University of Frankfurt. DP was supported by a grant from the Trond Mohn Foundation (project no. BFS2018REK01).


\begin{thebibliography}{99}

\bibitem{Mehtar-Tani:2013pia}
Y.~Mehtar-Tani, J.~G.~Milhano and K.~Tywoniuk,
Int. J. Mod. Phys. A \textbf{28} (2013), 1340013
doi:10.1142/S0217751X13400137
[arXiv:1302.2579 [hep-ph]].

\bibitem{Qin:2015srf}
G.~Y.~Qin and X.~N.~Wang,
Int. J. Mod. Phys. E \textbf{24} (2015) no.11, 1530014
doi:10.1142/S0218301315300143
[arXiv:1511.00790 [hep-ph]].

\bibitem{Bass:2008rv}
S.~A.~Bass, C.~Gale, A.~Majumder, C.~Nonaka, G.~Y.~Qin, T.~Renk and J.~Ruppert,
Phys. Rev. C \textbf{79} (2009), 024901
doi:10.1103/PhysRevC.79.024901
[arXiv:0808.0908 [nucl-th]].

\bibitem{Dorau:2019ozd}
P.~Dorau, J.~B.~Rose, D.~Pablos and H.~Elfner,
Phys. Rev. C \textbf{101} (2020) no.3, 035208
doi:10.1103/PhysRevC.101.035208
[arXiv:1910.07027 [nucl-th]].

\bibitem{Cassing:2003sb}
W.~Cassing, K.~Gallmeister and C.~Greiner,
Nucl. Phys. A \textbf{735} (2004), 277-299
doi:10.1016/j.nuclphysa.2004.01.127
[arXiv:hep-ph/0311358 [hep-ph]].

\bibitem{daSilva:2020cyn}
A.~V.~da Silva, W.~M.~Serenone, D.~Dobrigkeit Chinellato, J.~Takahashi and C.~Bierlich,
[arXiv:2002.10236 [hep-ph]].

\bibitem{Weil:2016zrk}
J.~Weil, V.~Steinberg, J.~Staudenmaier, L.~G.~Pang, D.~Oliinychenko, J.~Mohs, M.~Kretz, T.~Kehrenberg, A.~Goldschmidt, B.~Bäuchle, J.~Auvinen, M.~Attems and H.~Petersen,
Phys. Rev. C \textbf{94} (2016) no.5, 054905
doi:10.1103/PhysRevC.94.054905
[arXiv:1606.06642 [nucl-th]].


\bibitem{SMASH16}
  D.~Oliinychenko et al.,
  SMASH version 1.6, DOI: 10.5281/zenodo.3485108
  [https://doi.org/10.5281/zenodo.3485108].


\end{thebibliography}
\end{document}